# Depth profiling charge accumulation from a ferroelectric into a doped Mott insulator

M. Marinova[*,a], J. E. Rault[b], A. Gloter[*,a], S. Nemsak[c,d,e], G. K. Palsson[f,g], J.-P. Rueff[b], C. S. Fadley[c,d], C. Carrétéro[h], H. Yamada[h,i], K. March[a], V. Garcia[h], S. Fusil[h], A. Barthélémy[h], O. Stéphan[a], C. Colliex[a], M. Bibes[h]

[a] Laboratoire de Physique des Solides, CNRS UMR 8502, Université Paris Sud XI, 91405 Orsay, France

[b] Synchrotron-SOLEIL, BP 48, Saint-Aubin, F91192 Gif sur Yvette CEDEX, France

[c] Department of Physics, University of California Davis, Davis, CA 95616 USA

[d] Materials Sciences Division, Lawrence Berkeley National Laboratory, Berkeley, CA 94720 USA

[e] Peter Grünberg Institut PGI-6, Forschungszentrum Jülich, 52425 Jülich, Germany

[f] Department of Physics and Astronomy, Uppsala University, Box 516, SE-75120, Uppsala, Sweden

[g] Institut Laue-Langevin, 38000 Grenoble, France

[h] Unité Mixte de Physique CNRS/Thales,1 Avenue A. Fresnel, 91767 Palaiseau, France and Université Paris-Sud, 91405 Orsay, France

[i] National Institute of Advanced Industrial Science and Technology (AIST), JST, PRESTO, Tsukuba, Ibaraki 305-8562, Japan

KEYWORDS

Aberration-corrected STEM-EELS, Mott insulator, ferroelectricity, interface charge accumulation, interface physics, HAXPES

ABSTRACT

The electric field control of functional properties is a crucial goal in oxide-based electronics. Non-volatile switching between different resistivity or magnetic states in an oxide channel can be achieved through charge accumulation or depletion from an adjacent ferroelectric. However, the way in which charge distributes near the interface between the ferroelectric and the oxide remains poorly known, which limits our understanding of such switching effects. Here we use a first-of-a-kind combination of scanning transmission electron microscopy with electron energy loss spectroscopy, near-total-reflection hard X-ray photoemission spectroscopy, and ab-initio theory to address this issue. We achieve a direct, quantitative, atomic-scale characterization of the polarization-induced charge density changes at the interface between the ferroelectric $BiFeO_3$ and the doped Mott insulator $Ca_{1-x}Ce_xMnO_3$, thus providing insight on how interface-engineering can enhance these switching effects.

The ferroelectric control of functional properties in strongly correlated oxide nanostructures through electrostatic carrier density modulations has inspired intensive research efforts [1,2]. The field effect technique is particularly attractive for nanostructures consisting of complex oxides such as mixed-valence manganites, as these electron-correlated systems exhibit rich phase diagrams [3].

Hole-doped mixed-valence manganites with predominant $Mn^{3+}$ oxidation state have been extensively investigated [3–8]. In particular, recent works have been devoted on the interface between the half-metal $La_{1-x}Sr_xMnO_3$ (LSMO) manganite and ferroelectrics such as $Pb(Zr_xTi_{1-x})O_3$ [4,9], $BiFeO_3$ [10] and $BaTiO_3$ [11]. For example, reversible electrostatic accumulation and depletion of carriers has been obtained in $Pb(Zr_{0.2}Ti_{0.8})O_3/La_{0.8}Sr_{0.2}MnO_3$ [4] heterojunctions. An average charge transfer of 0.1 e/u.c. (unit cell) over 11 u.c., was observed depending on the ferroelectric polarization state [4]. This also resulted in a strong decrease in the magnetic moment at the Mn sites for the accumulation state. Spurgeon et al.[9] showed recently that local strain fluctuations also strongly influence the magnetization of the layers. Furthermore, a common report of these recent works is the non-symmetrical screening mechanism taking place in the LSMO layer when the polarisation state of the ferroelectric is reversed. For example, Chen et al.[11] reported very different orbital occupations in LSMO when the polarization direction in $BiFeO_3$ is reversed. This results in a strong distortion of the manganite for only one of the direction of the ferroelectric polarization. Kim et al.[10] suggested either an electronic or a predominately ionic screening mechanism in the LSMO depending on the $BiFeO_3$ polarization direction.

Studies on nanostructures based on electron-doped manganites like $Ca_{1-x}A_xMnO_3$ (A - La, Ce, Pr, Nd, Sm, Tb), with predominant $Mn^{4+}$ oxidation states, are much less abundant. In these manganites, a ferroelectrically induced interfacial phase transition can be more easily achieved by electrostatic doping, as lower amount of carrier-doping is necessary to introduce a metal-to-insulator transition [12].

Such ferroelectric control of functional properties is of particular importance in the case of Mott insulators, opening the possibility of the realisation of smaller or less energy consumption switches or memory devices. Recently, a strong electroresistance effect induced by the ferroelectric polarization switching has been demonstrated in heterostructures built with a ferroelectric $BiFeO_3$ and a Mott insulator based on the $Ca_{1-x}Ce_xMnO_3$ electron doped manganite [13,14].

In these recent reports [13,14], the resistivity change in the manganite channel induced by polarization switching in the ferroelectric $BiFeO_3$ was limited to one order of magnitude while changes of several orders were reported for thin films of $CaMnO_3$ tuned by different doping levels, strains or by electrolyte gating [12]. Typical electron injection of around 0.1 free electrons in the manganite channel were estimated by Hall measurements and associated with this resistivity change.

A charge distribution measurement on a unit cell scale at the Ca manganite/ferroelectric interface is thus required to understand, control and possibly enhance the magnitude of the resistivity switching effects. In the present study, a first-of-a-kind combination of scanning transmission electron microscopy (STEM) coupled to electron energy loss spectroscopy (EELS) [15–18], and grazing-incidence hard x-ray photoemission spectroscopy (HAXPES) in a near-total-reflection (NTR) condition have been used as site-sensitive valence probes to address the electronic and structural properties of $BiFeO_3$ and $Ca_{1-x}Ce_xMnO_3$ heterostructures (x=0, 2, 4 at.% nominal Ce concentrations) at the atomic scale. Our experimental results have also been compared to local density functional theory (DFT).

We have used these two novel and complementary methods to evaluate electron densities in manganite oxides. In the first, electron densities variation as small as ca. $0.02e^-$ are obtained with a unit cell resolution based on the analysis of the Mn-$L_3$ energy-loss near edge fine structures (ELNES). In the second, by tailoring the exciting hard x-ray wave field using the heterogeneous optical properties of the heterostructure, we acquire depth-resolved core-level photoemission spectra from each part of the heterostructure. Both methods independently reveal the charge accumulation profile at the $Ca_{1-x}Ce_xMnO_3$ interface with the downward polarized ferroelectric $BiFeO_3$. We want to point out the complementarity of both techniques, the first one probing unoccupied electronics states with high lateral spatial resolution in cross-section sample, the second one having depth profile resolution of occupied electronic state on the intact substrate. Such remarkable synergy might be applied to a large variety of interfacial systems.

Finally, as a guide to understanding the present results and being able to better tailor the properties of future nanostructures, we have used density functional theory (DFT) calculations to explore the influence of termination planes at the interfaces on the accumulation of carriers and the direction of the ferroelectric polarization.

A high-angle annular dark field (HAADF) image of a heterostructure of the type $YAlO_3$ (as a growth substrate)/$CaMnO_3$ (10 nm)/$Ca_{0.98}Ce_{0.02}MnO_3$ (10 nm)/$Ca_{0.96}Ce_{0.04}MnO_3$ (10 nm)/$BiFeO_3$ (5 nm) in the {110} pseudocubic projection is shown in Fig. 1(a). For reference, a HAADF image and the corresponding EELS for a bilayer heterostructure made of undoped and of 4 at % Ce-doped manganite layer ($YAlO_3$/$Ca_{0.96}Ce_{0.04}MnO_3$ (20 nm)/$CaMnO_3$ (10 nm)/$BiFeO_3$ (5 nm)), are given in respectively Figs. S1(a) and (b) of the Supporting Information. These multilayer systems allow the quantification of the spectroscopic signature of both undoped $CaMnO_3$ and nominally 2 and 4 at% Ce doped $Ca_{1-x}Ce_xMnO_3$, the latter being reported to be in a metallic state [12].

Fig. 1(b) shows the evolution of the c/a ratios of $BiFeO_3$ and $CaMnO_3$ as measured from the HAADF image in Fig. 1(c). The measured c/a ratio for $Ca_{1-x}Ce_xMnO_3$ is of the order of 1.02±0.03. In the vicinity to the $Ca_{1-x}Ce_xMnO_3/BiFeO_3$ interfaces the c/a ratio of the last $Ca_{1-x}Ce_xMnO_3$ unit cells slightly increases and in the $BiFeO_3$ it reaches within 3 unit cells the c/a ratio ≥1.25±0.03 which is indicative of the highly tetragonal $BiFeO_3$ phase [19]. The stabilization of this highly tetragonal $BiFeO_3$ phase (with ferroelectric polarization >100 $\mu C/cm^2$) is induced by the in-plane compressive strain of the $YAlO_3$ substrates [20].

Additionally the compressive epitaxial strain determines the electronic state of the $Ca_{1-x}Ce_xMnO_3$, i.e. it stabilizes a conductive state at lower doping concentrations [12]. The experimentally measured values for the c/a ratios coincide with the targeted compressively strained $Ca_{1-x}Ce_xMnO_3$ layers (the c/a ratio measured for the $YAlO_3$ substrate is 1.00±0.03), where the metallic character can be driven by doping concentrations of the formally tetravalent $Ce^{4+}$ ions between 1.5at% and 4.5at% or nominal electron doping between 0.03 e/u.c. and 0.09 e/u.c [12].

Such a narrow metallic region in the phase diagram of $Ca_{1-x}Ce_xMnO_3$ requires the characterization of any small doping variations and the related extra charges induced by the cation substitution. First, we used STEM-EELS to confirm the expected site and electronic occupation of Ce as shown in Fig. 1 (d)-(h). Although Ce is only present at the level of few percents, the STEM-EELS chemical maps confirm unambiguously that the Ce columns are correlated with the Ca columns. The presence of satellites at the higher energy side of the Ce-$M_{4,5}$ edges shown on Fig. 1h are typical fine structures observed in tetravalent cerium [21]. Therefore, the Ce substitution of a Ca atom is thus confirmed to provide two electrons per dopant atom in the manganite.

The electronic doping level in the manganite was also studied by the energy-loss near-edge fine structures. The transition metal $L_{3,2}$ edges reflect the changes of oxidation state upon doping as it can be seen from the Mn-$L_{3,2}$ edges of the undoped, 2at % and 4at % Ce doped layers shown in Fig. 2(a), where a small chemical shift is observed upon increasing the doping levels. In the same area, the local cationic concentration is measured by EELS at lower dispersion comprising Ca, O, Mn, Ce edges, as demonstrated by the EELS spectra in Fig. 2(b). The Mn L edges are dominated by two characteristic features and the intensity ratio between these two lines (the $L_3/L_2$ ratio) is often used to evaluate the Mn oxidation state from EELS spectra [22–24]. However, such a technique is strongly dependent on thickness and background extraction procedures and can impose difficulties when very small charge variations are studied. Other methods such as Mn-$L_3$ or O-K onset energies, energy differences between the O-K and Mn-$L_3$ edges or fitting procedures with reference spectra of known oxidation state have been used as well [24].

By contrast, in mixed-valence compounds, changes in the orbital occupancy are often evaluated by the Mn chemical shift in X-ray absorption spectroscopy (XAS) using Mn-$L_{3,2}$ [5,7,25–28] edges and this method can be employed in EELS studies to avoid artifacts linked to extraction procedures. In order to evaluate quantitatively the chemical shift observed in the Mn-$L_{3,2}$ edge in the EELS spectra, a calibration curve has been built on the basis of the energy shifts in a series of 2p XAS spectra of several mixed valence manganites which have structure similar to the $Ca_{1-x}Ce_xMnO_3$. The extracted chemical shifts, estimated as the barycenter of the Mn-$L_3$ from XAS references of the manganite series $Nd_{1-x}Ca_xMnO_3$ [7], $La_{1-x}Sr_xMnO_3$ [5], $CaMnO_x$ [24], with higher numbers of x values than our samples, are plotted as a function of doping concentration in Fig. 2(c). The shifts deduced from other manganite series like $La_{1-x}Ba_xMnO_3$ [25], $Pr_{1-x}Ca_xMnO_{3-\delta}$ [26,28], $Pr_{1-x}Sr_xMnO_{3-\delta}$ [26], $La_{1-x}Ca_xMnO_3$ [28] and $Tb_{1-x}Ca_xMnO_3$ [28] with a fewer number of doping concentrations, are included as well. One can clearly note the non-linear evolution of the Mn-$L_{3,2}$ shift versus the 3d orbital occupancy dependence. In particular the energy shift is stronger in the hole-doped regime and decreases on going into the electron doping regime. Indeed, it is noteworthy that in highly correlated systems like mixed-valence manganites, an anomalous spectral weight transfer is reported to appear upon doping [29] and the charge distributes differently among the oxygen and the metal at different doping levels. Hence from the dependence presented in Fig. 2(c) and the corresponding fitted curve, we can deduce that for the far-end of the electron doping regime a shift of 15 meV in the Mn-$L_{3,2}$ barycenter corresponds to introduction of 0.04 e/u.c which corresponds to a 2 at% Ce doping. The figure caption of fig. 2c needs to be clarified. A small clarification has been inserted in the figure caption.

In Fig 2(e) the evolution of the Mn-$L_3$ energy position across a nominal $Ca_{0.98}Ce_{0.02}MnO_3$ (10 nm) - $Ca_{0.96}Ce_{0.04}MnO_3$ (10 nm) – $BiFeO_3$ (5 nm) region of the heterostructure presented in the HAADF image (Fig. 2(d)) is shown. The Mn-$L_3$ position in Fig. 2(e) is given for each unit cell of the $Ca_{1-x}Ce_xMnO_3$ and it shows a shift of about 15 meV taking place at about 10 nm from the $BiFeO_3$. According to the calibration curve of Fig. 2(c), such 15 meV transitions correspond to typically 0.04 e/u.c which is expected from the nominal Ce dopant change occurring in these layers. Furthermore, the EELS measurement of the Ce/Ca concentration profile, shown in Fig. 2(f), confirms a change of the cerium content at similar depth. A similar correlation is found in the heterostructure $YAlO_3$/$Ca_{0.96}Ce_{0.04}MnO_3$ (20 nm)/$CaMnO_3$ (10 nm)/$BiFeO_3$ (5 nm), where a shift of 20 meV was deduced to correspond to 2.5 at% Ce or 0.05e/u.c. (See Supporting Information Fig. S2).

This shows that STEM-EELS can be used to quantify small electron doping changes with unit cell resolution if adequate references are used. These STEM-EELS data was then used together with the calibration curve in Fig. 2(c) to study the electron doping profile in the $Ca_{1-x}Ce_xMnO_3$ at the interface with the ferroelectric $BiFeO_3$ layer. The Mn-$L_3$ energy position profile in Fig. 2(e) exhibits a gradual shift of about 90 meV within 9-10 u.c. from the $Ca_{0.96}Ce_{0.04}MnO_3$/$BiFeO_3$ interface that is not due to Ce/Ca evolution. The excess charges per u.c. evaluated with respect to the nearby bulk-like and non-affected by the ferroelectric polarization $Ca_{0.96}Ce_{0.04}MnO_3$ region, are plotted in Fig. 2(g). Thus within 10 u.c. close to the $Ca_{0.96}Ce_{0.04}MnO_3$ /$BiFeO_3$ interface an integrated total charge accumulation of at least 0.65 $e^-$ is obtained by EELS. This value, measured spectroscopically in real space is smaller than the expected value of ca. 0.85$e^-$ for a downward oriented polarization (100 $\mu C/cm^2$) of $BiFeO_3$. However, the measured value is larger than the estimation from Hall effect measurements (ca. 0.1$e^-$) for the same $CaMnO_3/BiFeO_3$ interface [13], which suggests that the large excess charge located at the first two manganite cells (ca. 0.3 $e^-$) at the interface is not mobile. The reported metallic slab corresponds to the region of around 7 u.c. of manganite with typically below 0.05 $e^-$/u.c. This EELS data also confirms the rather large screening length inferred for this Mott insulator [13].

In addition, the element-specific chemical and electronic profile at the $Ca_{1-x}Ce_xMnO_3/BiFeO_3$ interface has been investigated by standing-wave (SW) hard x-ray photoemission spectroscopy (HAXPES), in which depth profiling of buried layers and interfaces is achieved in grazing incidence or in near total reflection (NTR) conditions. (More information on this recently developed method of synchrotron radiation characterization is given in the supporting information and the references therein). The SW-HAXPES technique has been successfully shown to provide depth-resolved chemical state and electronic structure for samples grown either as a superlattice multilayer mirror, as in the case of $La_{0.7}Sr_{0.3}MnO_3/SrTiO_3$[30], or on top of $Si/MoSi_2$ multilayer mirrors, as in the case of the Fe/MgO heterojunctions [31]. In this study we show that the multilayer mirror is not necessary, by using x-ray wave interference effects in NTR conditions for a bilayer system.

Fig. 3(a) first shows the simulated x-ray intensity (as square of its electric field, E, for h$\nu$=2800 eV) as a function of the incidence angle, $\theta$, over the NTR range, including total reflection for zero incidence angle (see Supporting Information for the calculation details) [32]. This reveals that, for increasing incidence angle $\theta$, the x-rays penetrate deeper into the bilayer. A schematic representation of the bilayer $BiFeO_3/Ca_{1-x}Ce_xMnO_3$ heterostructure is given in the Fig. 3(b). In Fig. 3(c) and (d) are shown the Ca 2p core-level spectra at incidence angles θ = 0.69° (more interface sensitive) and 1.5° (more bulk sensitive), respectively. Each member of the spin-orbit-split doublet shows two components with a binding energy (BE) difference Δ BE=0.70 eV. This indicates that the Ca atoms in the $Ca_{1-x}Ce_xMnO_3$ layer are in two distinct chemical, structural or electronic potential environments. The spectral weight of the high-binding-energy (HBE) component decreases relative to the low-binding-energy (LBE) one when going from θ = 0.69° to 1.5°. In Fig. 3(a) one can see that at θ = 0.5° the x-rays field barely reaches the top of the $Ca_{1-x}Ce_xMnO_3$ layer while at θ = 1.5° the intensity is significant well down into the substrate. Therefore we can safely allocate the HBE component to an interface region and LBE component to the bulk of the $Ca_{1-x}Ce_xMnO_3$. More quantitative insight on the depth distribution of all the chemical components present in the structure can be inferred by the analysis of the full set of NTR curves.

Fig. 3(e) shows the experimental NTR curves (blank circles) for C 1s, Bi 4f, interface Ca 2p and bulk Ca 2p core-levels. As expected from the electric-field calculations (Fig. 3(a)), the NTR curves clearly shows significant intensity oscillations as a function of the photon incidence angle (e.g. ≈ 20% above the total-reflection falloff and located at 0.8° for the Bi 4f core-level). The oscillations are also phase-shifted when going from one layer to another (e.g. 0.20° when going from $BiFeO_3$ to bulk $Ca_{1-x}Ce_xMnO_3$), which demonstrates the incidence-angle dependence of the probing depth provided by the SW phase information in the field profile. To obtain a quantitative depth profile of the structure we compare the experimental NTR curves to calculations done with the Yang X-Ray Optics (YXRO) software for photoemission [32]. The optimal structure that best describes the data consists of a top CO contamination layer (10 Å), a $BiFeO_3$ layer (with thickness of 42 Å) and $Ca_{1-x}Ce_xMnO_3$ layer divided in an interface (10 Å) and a bulk component (210 Å). These optimal thicknesses determined from the calculation are summarized in Fig. 3(b). The best fit to the experimental curves (solid lines in Fig. 3(e)) is thus obtained for interface $Ca_{1-x}Ce_xMnO_3$ layer with thickness $10 \pm 2\text{-}3$ Å (i.e. 2.5 u.c.). This result is in good agreement with the results obtained by STEM-EELS, which show the highest charge concentration over the first two unit cells close to the interface with the ferroelectric $BiFeO_3$.

However, in most compounds, electron-doping induces a shift towards lower BEs. For instance in $BaTiO_3$, electron doping induced by oxygen vacancies leads to the appearance of a $Ti^{3+}$ peak at lower BE than the $Ti^{4+}$ peak [33]. Nevertheless, in $ABO_3$ compounds, one would expect a different behavior between the covalent $BO_2$ and the ionic AO layers. Vanacore et al. showed this is indeed the case in $SrTiO_3$ single crystals [34]. The surface-related HBE peak of Sr 3d is due to a lower coordination, or equivalently a higher electron occupation, at surface cation sites. The same behavior has been observed in $BaTiO_3$ [35]. Van der Heide showed that Ca, which is in the same periodic table column as Sr and Ba, behaves similarly i.e. that a higher electrons occupation induces higher-binding-energy for Ca core-levels [36]. Finally, Taguchi and Shimada showed that oxygen vacancies induce a HBE peak in $CaMnO_3$ [37]. All of this suggests that the HBE component we see in Fig. 3(c) is indeed due to a local electron doping of the $Ca_{1-x}Ce_xMnO_3$ layer. The NTR-HAXPES results are thus in agreement with the charge accumulation at the $Ca_{1-x}Ce_xMnO_3/BiFeO_3$ interface obtained by the STEM-EELS.

A charge accumulation at the ferroelectric/manganite interface can be understood in the view of electrostatic electron doping due to the ferroelectric polarization [1,4]. In this aspect, the B-site atomic displacement with respect to the A site, associated to the ferroelectric polarization, can be measured by STEM-HAADF (oxygen atomic positions are not legible in such images) and is plotted in Fig. 1(b) for each unit cell. The average displacement of the Fe ions, in the region of $BiFeO_3$ where the c/a ratio has reached values higher than $1.25\pm0.03$, was determined to be 40 pm $\pm$ 15 pm along the [001] direction. Such a value is similar to a bulk like displacement (see $BiFeO_3$ model structure in figure 1(i)) and thus corresponds to polarization of the order of P~100 µC/cm$^2$ [38] pointing towards the BiFeO$_3$/Ca$_{1-x}$Ce$_x$MnO$_3$ interface [38]. This value is also in agreement with piezoresponce force microscopy measurements [13,14]. Nevertheless, the observed amount of electrons doping in the CaMnO$_3$ as determined by EELS in this manuscript and as reported from the Hall effect [13] is lower than expected from the screening of $P \approx 100$ µC/cm$^2$.

Structural rearrangement at this interface can also screen the polarization of the ferroelectric layer and thus alter the numbers of electrons induced in the manganite. Combining Ca, Fe and Mn L$_{3,2}$ signals and the HAADF intensity profile (Figs. 4(a)-(b)), the identification of the preferential termination planes at the BiFeO$_3$/Ca$_{1-x}$Ce$_x$MnO$_3$ interface has been achieved. The BiFeO$_3$ is found to be B-cation terminated (i.e. FeO$_2$ terminated), giving at the interface the (BiO)$^{+1}$-(FeO$_2$)$^{-1}$-(CaO)$^0$-(MnO$_2$)$^0$ sequence, resulting in a polar discontinuity. Such polar discontinuity suggests introduction of holes as discussed previously for systems like LaAlO$_3$-SrTiO$_3$ [17], La$_{1-x}$A$_x$MnO$_3$-SrTiO$_3$ [39], La$_{0.6}$Sr$_{0.4}$MnO$_3$/Nb:SrTiO$_3$ [18,40].

In order to evaluate the role of the interface termination in the charge distribution in CaMnO$_3$, DFT modeling of both interfacial terminations, *i.e.*, when the ferroelectric vector points towards the (FeO$_2$)$^{-1}$-(BiO)$^{+1}$- (MnO$_2$)$^0$ -(CaO)$^0$ or the (BiO)$^{+1}$-(FeO$_2$)$^{-1}$-(CaO)$^0$-(MnO$_2$)$^0$ interfaces has been carried out. Both relaxed models in the case of 6 and 7 unit cells of BiFeO$_3$ and of CaMnO$_3$ can be seen in Fig. 4c and 4d. The central BiFeO$_3$ unit cells have been kept fixed with a structure identical as in the bulk while the other BiFeO$_3$ and CaMnO$_3$ atoms have been allowed to relax. The LSDA+U approximation with U=5eV on the iron site has been used in order to introduce stronger electronic correlation in the BiFeO$_3$ layer and to maintain an insulating barrier. After structural relaxation, the case where the ferroelectric polarization points towards the (FeO$_2$)$^{-1}$-(CaO)$^0$-terminated interface is found lower in energy of around 1.5 eV. This is in accordance with the previous calculation of Yu et al [41] who found in related BiFeO$_3$-(La,Sr)MnO$_3$ interfaces a preferential energy when the polarization vector of BiFeO$_3$ points towards the (La,Sr)O terminated electrode. This is also in agreement with the downward polarization measured by STEM-HAADF for the (FeO$_2$)$^{-1}$-(CaO)$^0$-terminated interface (Fig. 1b).

For this termination, the ferroelectric displacements in the BiFeO$_3$ layers obtained from the DFT relaxations are larger and an average projected Fe-O$_{apical}$ displacements of 64 pm is found. In the less energetically favored case where the BiFeO$_3$ polarization points toward the (BiO)$^{+1}$-(MnO$_2$)$^0$-terminated interface, a smaller average Fe-O$_{apical}$ atomic displacements of 55 pm is found in the BiFeO$_3$.

The calculations also indicate a different reconstruction of the manganite unit cell next to the interface that is mostly governed by the interface termination. An average atomic displacement of the Mn cation with respect to the A-site of around 24pm is obtained in the case of (BiO)$^{+1}$-(MnO$_2$)$^0$-terminated interface, while it is of only 10 pm for the (FeO$_2$)$^{-1}$-(CaO)$^0$-terminated interface (in agreement with the STEM-HAADF measurements of the figure 1b). These distortions are also modulated by the direction of the ferroelectric polarization (as previously reported in the case of a BaTiO$_3$-LSMO interface[11]), but this effect is weaker and of only 5-10 pm according to the *ab-initio* calculation. The stronger distortion for one interface termination can be easily understood since only the (BiO)$^{+1}$-(MnO$_2$)$^0$-terminated interface results in a asymmetric chemical environment with BiO layer on one side and CaO layer on the other side, and the presence of this Bi 6s$^2$ states strongly favors the distortion.

We investigated the influence of the termination planes on the charge induced in the d orbitals of the Mn, which can be related to the STEM-EELS Mn-L$_3$ measurement and the interface shift in the NTR-HAXPES results. For both terminations, the charges in the manganite electrode where the polarization points out (the Mn$_{Bottom}$ side in the Fig. 4) are very similar with the one in the center of the CaMnO$_3$ or in the calculated bulk. This indicates that the injection of holes in the CaMnO$_3$ electrodes is not a favorable process and it can also be seen by the similarity of the density of states of the Mn d orbitals for the Mn$_{Bottom}$ and the Mn in the central position. The situation is very different for the electrode located where the BiFeO$_3$ polarization points in (the Mn$_{Top}$ side). In the case of the polarization pointing to the (BiO)$^{+1}$-(MnO$_2$)$^0$-terminated interface, there are respectively 0.15, 0.03 and 0.01 electrons which are injected in the e$_g$ orbital of the first three Mn atoms. In contrast, less electrons are injected for the (FeO$_2$)$^{-1}$-(CaO)$^0$-terminated interface where only 0.04 and 0.02 additional electrons are reported for the two first Mn atoms. The difference can also be seen in the density of states of the Mn d orbitals. In the case of the (BiO)$^{+1}$-(MnO$_2$)$^0$-terminated interface the shift to lower energy of the band is higher of around 0.7 eV, the shape of the density of states being strongly altered, lowering e$_g$ states below the Fermi level. This confirms the role of the interface polar discontinuity to control the native ferroelectric polarization but also the numbers of injected charge in the manganite electrodes. Indeed, although the BiFeO$_3$ layers in the case of (FeO$_2$)$^{-1}$-(CaO)$^0$-terminated interface have a stronger ferroelectric polarization, the numbers of electrons in the manganite is lower.

Both STEM-EELS and DFT results thus explain why despite the observed large ferroelectric behavior in the BiFeO$_3$ layers, the number of injected electrons stays rather low, due to the polar discontinuity of the (FeO$_2$)$^{-1}$-(CaO)$^0$ - terminated interface. Nevertheless, the total amount of electrons observed by STEM-EELS confirms that several nanometers of Ca$_{1-x}$Ce$_x$MnO$_3$ next to the BiFeO$_3$ are in the electron doped and strained regime where a lower resistivity is expected. The NTR-HAXPES results support this conclusion. Poling the polarization of the BiFeO$_3$ away from the interface with the Ca$_{1-x}$Ce$_x$MnO$_3$ results then in a much higher resistivity as reported for this device.

Similar role of the plane termination might be partially at the origin of the recently reported asymmetric behavior of the EELS Mn L$_{2,3}$ energy shift in LSMO for the two different polarization directions of Pb(Zr$_x$Ti$_{1-x}$)O$_3$ [9]. It may also play a role in the difference of screening mechanism of the BaTiO$_3$ polarization in LSMO where purely electronic phenomena is suggested for one direction while oxygen vacancies are proposed for the other direction [10].

In summary the present study reports on the structure and the electronic properties of heterostructures between the electron-doped Mott insulator Ca$_{1-x}$Ce$_x$MnO$_3$ and the tetragonal BiFeO$_3$ ferroelectric gate. A method to evaluate small electron density modulation at the unit cell scale using STEM-EELS datasets by quantifying the Mn-L$_3$ chemical shifts has been introduced. Thus the accumulation of charge densities at the manganite interface with the adjacent ferroelectric has been revealed. NTR-HAXPES has also been used as a complementary method to derive the specimen morphology, including depth-resolved electronic structures, yielding results in excellent agreement with STEM-EELS. Finally, ab-initio calculations have revealed the importance of interfacial polar discontinuities in determining both the inherent direction of the ferroelectric polarization and the charge carrier density accumulated at this BiFeO$_3$/Ca$_{1-x}$Ce$_x$MnO$_3$ interface. Thus, the present study on the charge profile at the interface between a Mott insulator and a ferroelectric contributes to the understanding of interface phenomena in nanostructures between ferroelectrics and correlated oxides, and more specifically to the understanding of the electroresistance switching behavior of oxide devices based on such nanostructures.

The combination of depth resolved photoemission and laterally resolved EELS, as well as NTR-HAPXES, reported here, can also be extended to the study of other interfacial systems. As a future suggestion, the analysis of angle-resolved photoemission (ARPES) in NTR-HAXPES, should provide direct access to the depth-resolved interface band structure, as demonstrated recently in soft x-ray standing-wave ARPES [42].

FIGURES

**Figure 1.** (a) A HAADF image of a heterostructure of the form $YAlO_3/CaMnO_3$ (10 nm)/ $Ca_{0.98}Ce_{0.02}MnO_3$ (10 nm)/$Ca_{0.96}Ce_{0.04}MnO_3$ (10 nm)/$BiFeO_3$ (5 nm) in the {110} pseudocubic projection. (b) Evolution of the c/a ratio (green dots) and of the displacement of the B site (Fe or Mn ions) with respect to the A site (Bi or Ca ions) in the vicinity of the $Ca_{1-x}Ce_xMnO_3/BiFeO_3$ interface (red dots) as measured from the HAADF image (blue dots) shown in panel (c). The error bars correspond to an interval around the plotted mean values comprising 80% of the individual measurements. The change of HAADF intensity across the interface for one atomic row is given for a guideline in (b) as a dotted blue line. (d)-(f) Ca-$L_{3,2}$, Mn-$L_{3,2}$ and Ce-$M_{5,4}$ EELS maps. (g) on-line HAADF image. (h) ELNES spectra of Ce-$M_{5,4}$ edge confirming that Ce is in a tetravalent oxidation state. (i) Atomic model of bulk tetragonal $BiFeO_3$. The Fe ions displacement, $\delta_{Fe}$, is measured along the [001] direction in the $\{110\}_{pc}$ projection. Dark violet/black circles indicated Bi ions, blue/black circles – Fe ions and red/black indicate O ions.

**Figure 2.** (a) ELNES spectra of Mn-$L_{3,2}$ edges of the undoped, 2 at% Ce and 4 at% Ce doped layers, after background subtraction, where a small chemical shift and a growing low-energy shoulder are evident upon increasing $Ce^{4+}$ substitution is evident upon increasing $Ce^{4+}$ substitution. (b) EELS spectra for undoped, 2 at% Ce and 4 at% Ce doped layers, in the energy range between 350 and 950 eV, where Ca-$L_{3,2}$, O-K, Mn-$L_{3,2}$ and Ce-$M_{5,4}$ edges are seen. (c) Chemical shift vs dopant concentration (or oxygen concentration in the case of $CaMnO_x$ [27]) for different mixed valence manganites as extracted from XAS data and corresponding to, as follows: $La_{1-x}Sr_xMnO_3$[5] - white/red circles, $Nd_{1-x}Ca_xMnO_3$[7] - white/blue

circles, $La_{1-x}Ba_xMnO_3$[25] - grey full triangles, $Pr_{1-x}Ca_xMnO_{3-\delta}$[26] - full red circles and $Pr_{1-x}Ca_xMnO_3$[28] - black "plus" signs, $Pr_{1-x}Sr_xMnO_{3-\delta}$[26] - grey plus signs, $CaMnO_x$[27] - white/black circles, $La_{1-x}Ca_xMnO_3$[28] - back full circles $Tb_{1-x}Ca_xMnO_3$[28] - red full triangles. (d) Extract from the upper part of the HAADF image in Figure 1(a). The image is given as a guideline. (e) Mn-$L_3$ peak position extracted from ELNES spectra unit cell by unit cell. (f) Ce concentration profile for the same heterostructure. (g) Excess charge per unit cell at the interface between the $Ca_{0.96}Ce_{0.04}MnO_3$ and the $BiFeO_3$, as derived from (c). The error bars in panel (e) and (f) correspond to 80% confidence level.

**Figure 3.** (a) Calculation of the depth-resolved electric field strength $|E^2|$ as a function of depth and photon incidence angle. This demonstrates that changing the incidence angle permits selectively probing different depth. The oscillatory features represent standing-wave formation. (b) Schematic of the $BiFeO_3/Ca_{1-x}Ce_xMnO_3//YAlO_3$ heterostructure with the thicknesses of the layers obtained after fitting the experimental rocking curves by using the YRXO software[32] also indicated. Ca 2p core-level spectrum taken at incidence angles of (c) θ=0.69° and (d) θ=1.5°. The decrease of the HBE component when going from incidence angle 0.69° (interface) to 1.5° (bulk) suggests that this component comes belongs to the interface layer. (e) Experimental (blank circles) rocking curves for C 1s, Bi 4f, interface Ca 2p and bulk Ca 2p core-levels, overlaid with the theoretical curves [32].

**Figure 4.** (a) On-line HAADF image (b) Averaged normalized EELS intensity profiles of the Fe-$L_{3,2}$, Ca-$L_{3,2}$ and Mn-$L_{3,2}$ signals, respectively, at the $CaMnO_3/BiFeO_3$ interface. (c) and (d) Relaxed models of the $CaMnO_3/BiFeO_3$ interface with 6 and 7 unit cells of $BiFeO_3$ and $CaMnO_3$ and termination planes $(FeO_2)^{-1}$-$(BiO)^{+1}$-$(MnO_2)^0$-$(CaO)^0$ and $(BiO)^{+1}$-$(FeO_2)^{-1}$-$(CaO)^0$-$(MnO_2)^0$, respectively. (e) and (f) Densities of states of the Mn d orbitals for interfaces with termination planes $(FeO_2)^{-1}$-$(BiO)^{+1}$-$(MnO_2)^0$-$(CaO)^0$ and $(BiO)^{+1}$-$(FeO_2)^{-1}$-$(CaO)^0$-

$(MnO_2)^0$ respectively. $Mn_{Bottom}$ indicates the Mn cation where the ferrolectric polarization points outside the manganite and $Mn_{Top}$ when the ferroelectric polarization points towards the manganite. $Mn_{Central}$ indicates a Mn cation in the central $CaMnO_3$.

ASSOCIATED CONTENT

**Supporting Information**. Details on the film growth, STEM-EELS and the ab-initio modeling. Information on the NTR-HAXPES method and the calculation details used for the interpretation of the experimental results. Discussion of the charge modulation within the bilayer $BiFeO_3/Ca_{1-x}Ce_xMnO_3/YAlO_3$ (x=0,4 at% Ce) heterostructure as deduced from the Mn-$L_3$ shift. This material is available free of charge via the Internet at http://pubs.acs.org.

AUTHOR INFORMATION

**Corresponding Author**

*E-mail: (A.G.) alexandre.gloter@u-psud.fr

*E-mail: (M.M.) E-mail: maya.marinova@univ-lille1.fr

ACKNOWLEDGEMENT

This work was supported by the French Agence Nationale de la Recherche NOMILOPS project (ANR-11-BS10-0016) and 7th framework EU program ESTEEM2 (grant agreement 312483). We acknowledge financial support from the European Research Council (ERC Advanced Grant FEMMES, No. 267579) and ERC Consolidator Grant MINT, No.615759. JER, JPR, and CSF acknowledge the support of a public grant from the "Laboratoire d'Excellence Physics Atom Light Matter" (LabEx PALM) overseen by the ANR as part of the "Investissements d'Avenir" program (reference: ANR-10-LABX-0039). SN and CSF acknowledge partial support from the Army Research Office, under the Multidisciplinary University Research Initiative Grant W911-NF-09-1-0398. CSF is also supported for salary from the Director, Office of Science, Office of Basic Energy Sciences, the Materials Sciences and Engineering Division under the Department of Energy Contract No. DE-AC02-05CH11231), and Beamline 9.3.1 of the Advanced Light Source, at which some of the HAXPES was performed, is also supported under the same contract.

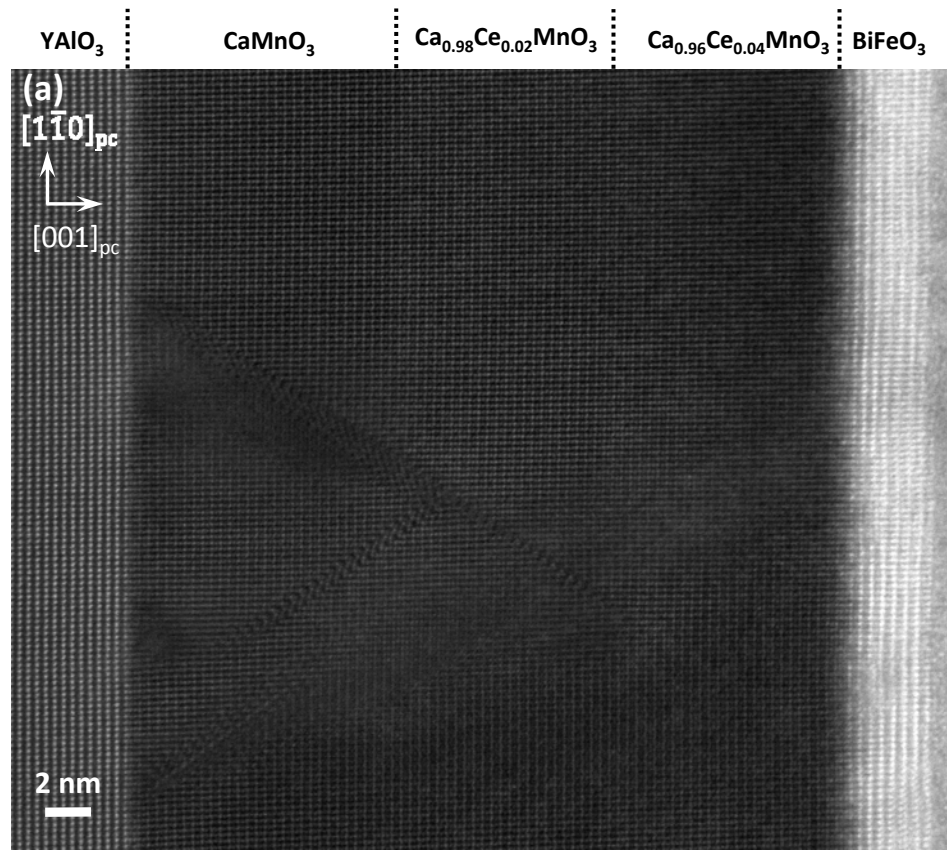
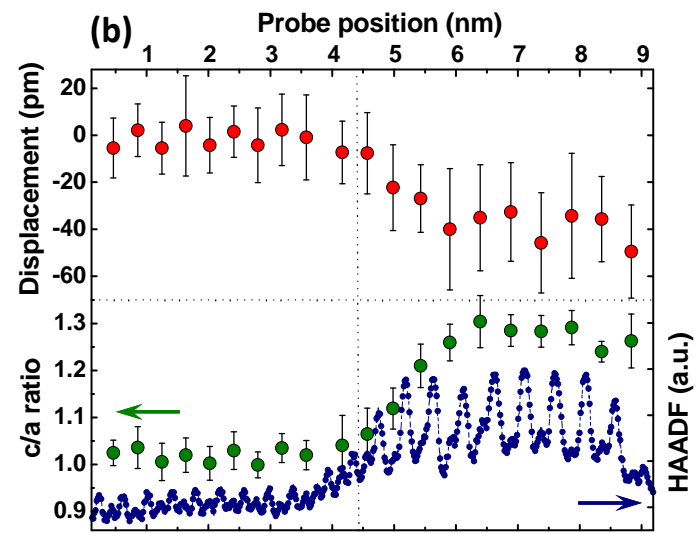
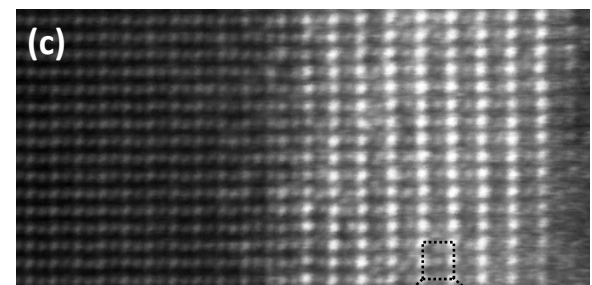
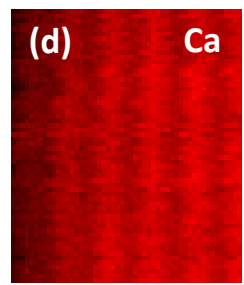 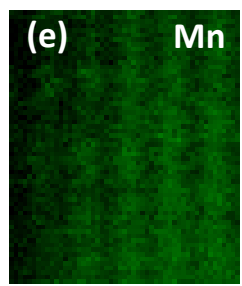
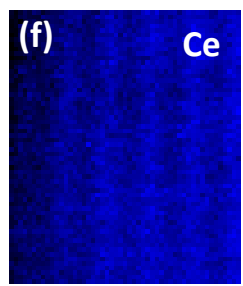 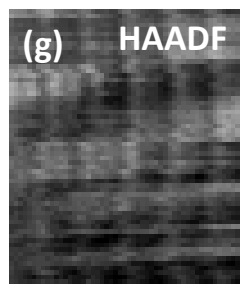
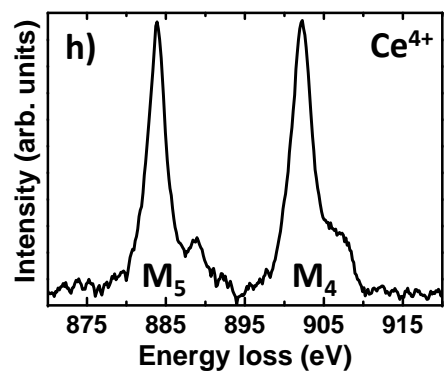
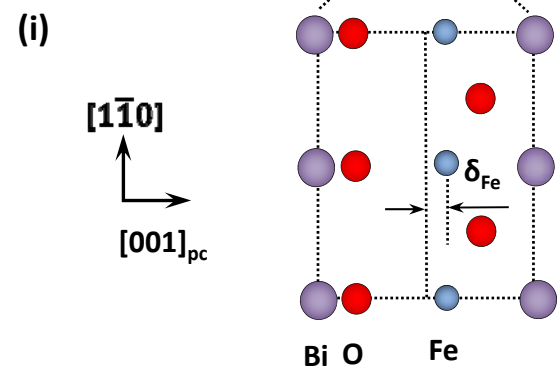

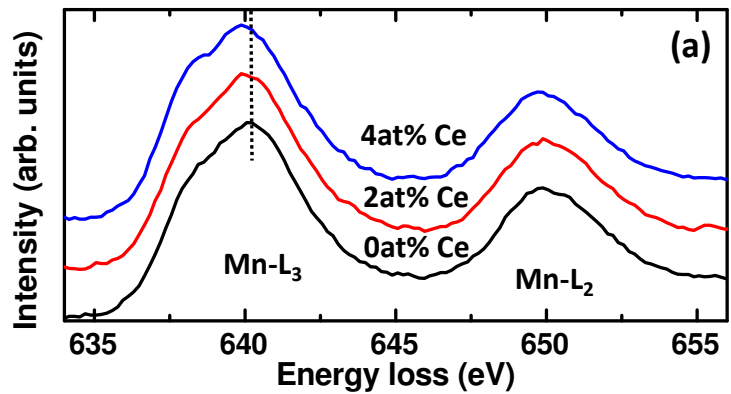
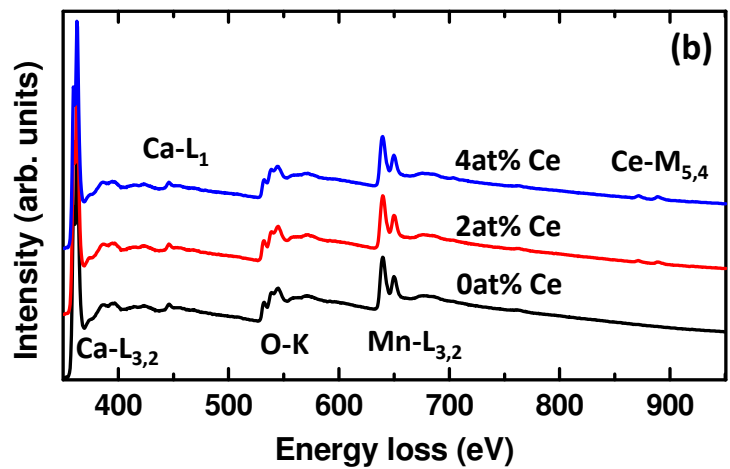
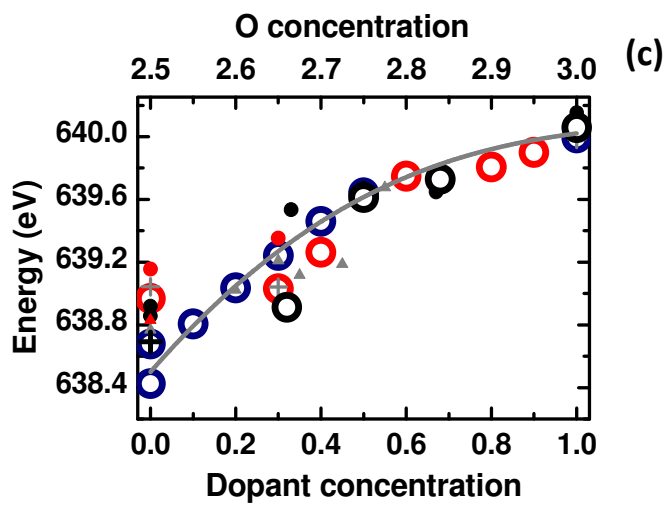
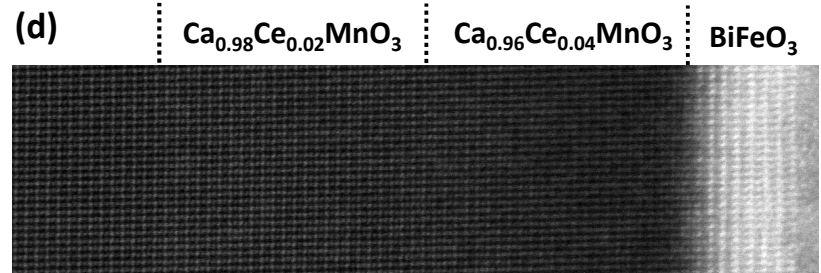
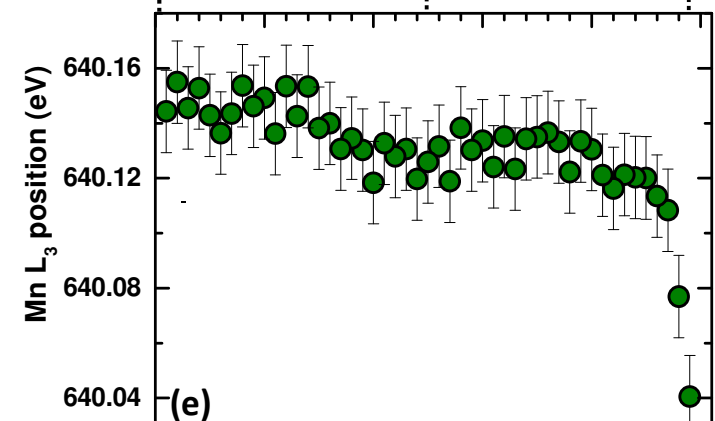
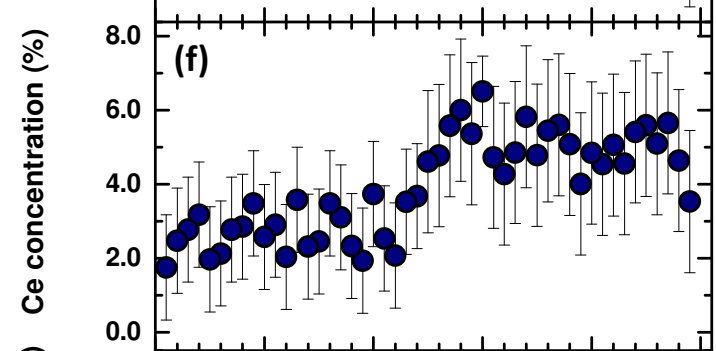
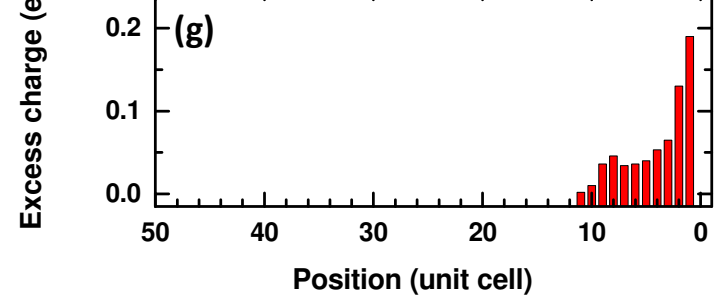

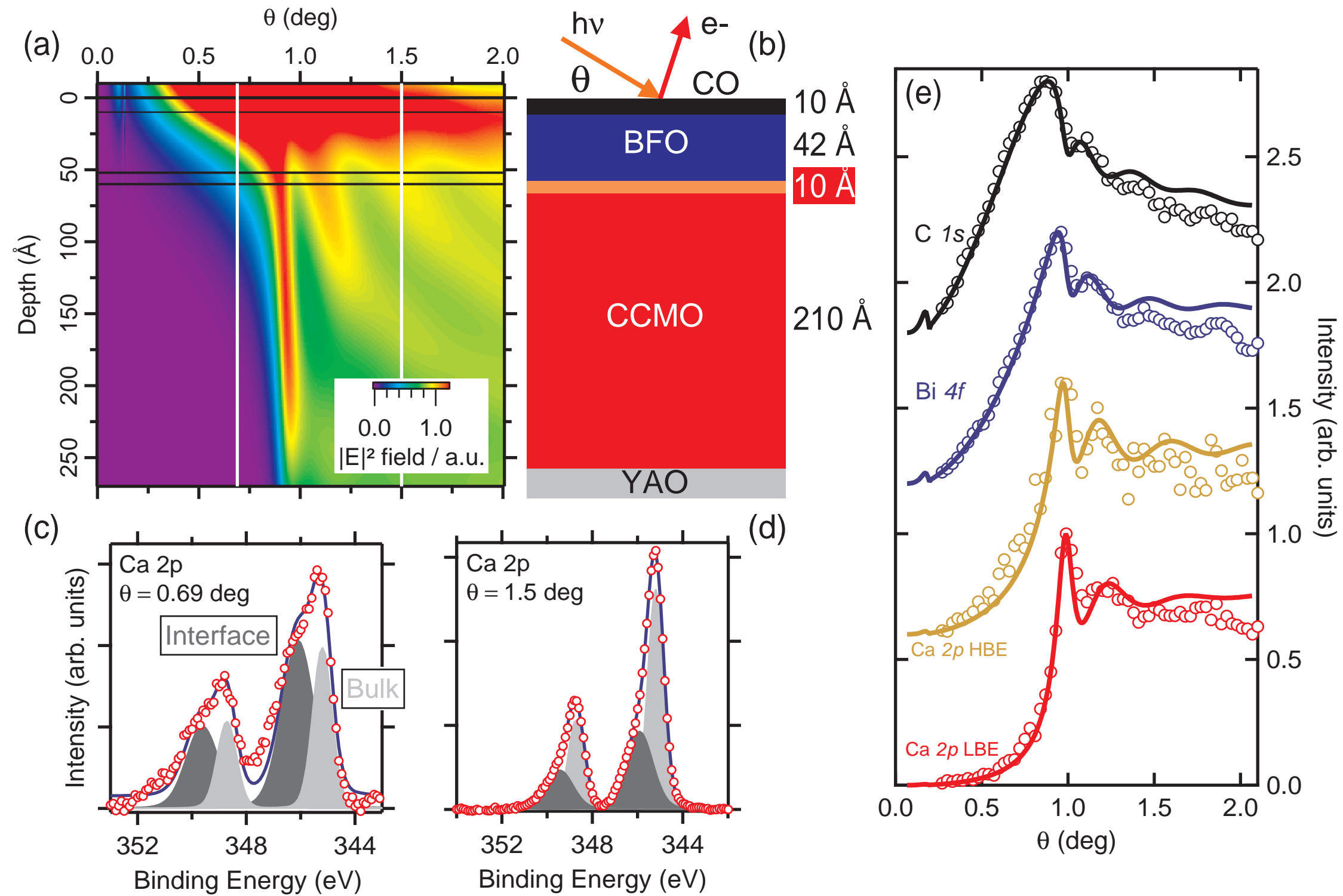

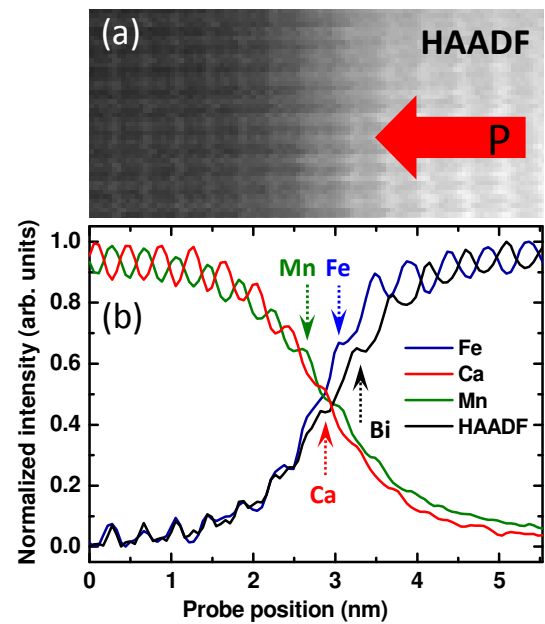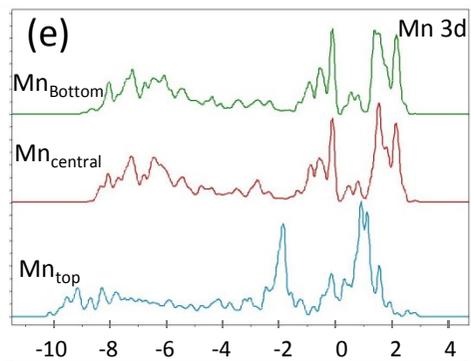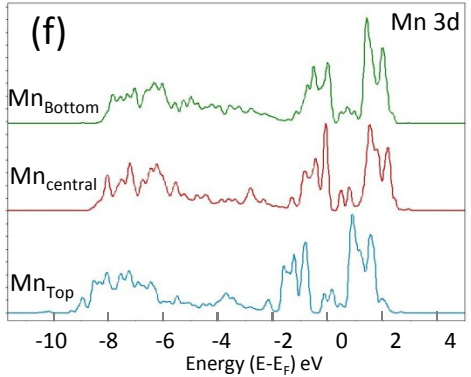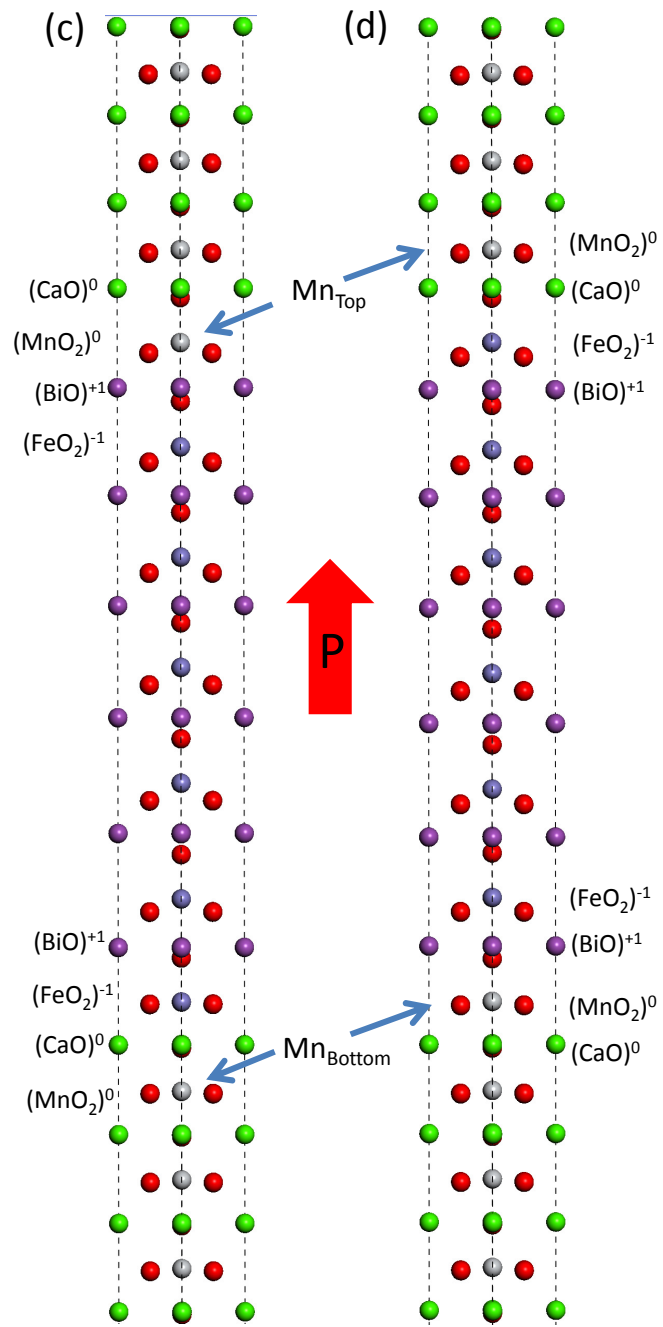